# A Dynamic ID-based Remote User Authentication Scheme

Manik Lal Das, Ashutosh Saxena, and Ved P. Gulati

**Abstract** — *Password-based authentication schemes are the most widely used techniques for remote user authentication. Many static ID-based remote user authentication schemes both with and without smart cards have been proposed. Most of the schemes do not allow the users to choose and change their passwords, and maintain a verifier table to verify the validity of the user login. In this paper we present a dynamic ID-based remote user authentication scheme using smart cards. Our scheme allows the users to choose and change their passwords freely, and do not maintain any verifier table. The scheme is secure against ID-theft, and can resist the reply attacks, forgery attacks, guessing attacks, insider attacks and stolen verifier attacks.[1]*

## I. INTRODUCTION

Password-based authentication schemes are the most widely used methods for remote user authentication. The resources distributed among the hosts are shared across the network in the form of network services provided by the remote systems. Before providing such services, the remote system should have ability to authenticate the users. Otherwise, an adversary could impersonate a legal user login to get access of the system. Existing password-based authentication schemes [1], [4]-[8], [10], [11] can be categorized into two types - one uses weak-password and the other uses strong-password. The weak-password authentication scheme is based on public-key cryptographic techniques (e.g. ElGamal's scheme [3]) and has the advantage that the remote system does not need to keep a verifier table to verify the validity of the user login. Though, weak-password is easy to memorize; however, weak-password authentication schemes lead heavy computational load to the whole application system because of using public-key cryptographic techniques. In contrast, the computational load of most strong-password authentication schemes is lighter because of using only simple operations, e.g., one-way hash function [9] and exclusive-OR operation. The strong-password authentication schemes have another advantage over weak-password authentication schemes that their implementations are easier and with less cost. However, a strong-password is difficult to memorize. Additionally, the strong-password authentication schemes suffer from stolen verifier attacks and guessing attacks. Several schemes and improvements [1], [2], [4]-[8], [10], [11] have been proposed, but these schemes are based on static login ID. There are numerous applications where static login ID leaks partial information about the user's login message to the adversary. The adversary could intercept the login ID and later try to manipulate it with other intercepted parameters to forge the login ID. One of the applications is digital library, where the subscriber needs to read and/or download the articles. Therefore, employing a dynamic ID for each login can avoid the risk of ID-theft. In this paper, we propose a dynamic ID-based remote user authentication scheme using smart cards. The proposed scheme allows the users to choose and change their password freely. The scheme is protected from ID-theft and the security of the scheme is based on a one-way hash function [9]. Moreover, the scheme does not maintain any verifier table, which strengthens the scheme against stolen verifier attacks. We shall analyze the security of the scheme and show that the scheme can resist other attacks, namely, the replay attacks, forgery attacks, guessing attacks and insider attacks.

## II. THE PROPOSED SCHEME

The scheme is composed of two parts, namely, the registration phase and the authentication phase. The registration phase is performed only once, and the authentication phase is executed every time the user logs into the system. The notations used throughout this paper are as follows:

| | |
|---|---|
| $U$ | the user. |
| $PW$ | the password of $U$. |
| $S$ | the remote system. |
| $h(.)$ | a one-way hash function. |
| $\oplus$ | bitwise XOR operation. |
| $A \Rightarrow B: M$ | $A$ sends $M$ to $B$ through a secure channel. |
| $A \rightarrow B: M$ | $A$ sends $M$ to $B$ through a common channel. |

### A. Registration Phase

This phase is invoked whenever a user $U_i$ registers to the remote system. The user chooses password $PW_i$ and submits it to the remote system. Upon receiving the registration request, the remote system performs the following steps:

1. Computes a nonce $N_i = h(PW_i) \oplus h(x)$, where $x$ is a secret key of the remote system.
2. Personalizes the smart card with the parameters $h(.)$, $N_i$ and $y$; where $y$ is a remote system's secret number stored in each registered user's smart card.
3. $S \Rightarrow U_i$: $PW_i$ and smart card.

### B. Authentication Phase

This phase is invoked whenever a user wants to login the remote system. The phase is further divided into two parts, namely, the login phase and the verification phase. In the login

[1]The work was supported in part by the Ministry of Communications and Information Technology, Govt. of India, under the grant no. DIT/R&D/Coord/1(6)/2003.

The authors are with Institute for Development and Research in Banking Technology, Castle Hills, Road No.1, Masab Tank, Hyderabad, INDIA (email: {mldas, asaxena, vpgulati}@idrbt.ac.in).

phase, U will send a login message to S. The login message contains a dynamic ID, called $CID$, which is dependent on the user password and remote system's secret parameters. After successful verification of the login message, the remote system allows the user to access the system. The login phase and the verification phase work as follows:

*Login Phase:* The user $U_i$ inserts his smart card to the card reader of a terminal, and keys his password $PW_i$. Then, the smart card will perform the following operations:
1. Computes $CID_i = h(PW_i) \oplus h(N_i \oplus y \oplus T)$, where $T$ is the current date and time of $U_i$'s system.
2. Computes $B_i = h(CID_i \oplus h(PW_i))$.
3. Computes $C_i = h(T \oplus N_i \oplus B_i \oplus y)$.
4. $U_i \to S$: $CID_i, N_i, C_i, T$.

*Verification Phase:* Upon receiving the login message ($CID_i, N_i, C_i, T$) at the time $T^*$, the remote system authenticates the user $U_i$ with the following steps:
1. Verify the validity of the time interval between $T$ and $T^*$. If $(T^* - T) \geq \Delta T$, where $\Delta T$ denotes the expected valid time interval for transmission delay, then the remote system rejects the login request.
2. Computes $h(PW_i) = CID_i \oplus h(N_i \oplus y \oplus T)$.
3. Computes $B_i = h(CID_i \oplus h(PW_i))$. Thereafter, checks whether $C_i = h(T \oplus N_i \oplus B_i \oplus y)$. If it holds, the remote system accepts the login request. Otherwise, rejects the login request and terminates the operation.

### C. Password Change Phase

This phase is invoked whenever the user $U_i$ wants to change his password. He can easily change his password without taking any assistance from the remote system. The phase works as follows:
1. The user $U_i$ inserts the smart card into the smart card reader of a terminal. He submits the password $PW_i$ and requests to change the password.
2. Then $U_i$ chooses new password $PW_{i*}$.
3. The smart card computes
   $N_{i*} = N_i \oplus h(PW_i) \oplus h(PW_{i*})$, which yields
   $h(PW_{i*}) \oplus h(x)$.
4. The nonce $N_i$ will be replaced with $N_{i*}$. The password has been changed with the new password $PW_{i*}$ and terminates the operation.

## III. SECURITY ANALYSIS

In the following, we analyze the security of our scheme:
1. A replay attack (replaying the intercepted login message) cannot work because it will fail the Step 1 of verification phase for the time interval $(T^* - T) \geq \Delta T$.

2. An adversary cannot impersonate a legal login even he intercepts $CID_i, N_i, C_i$ and $T$. To forge a valid login, the adversary needs to know two secret parameters $B_i (= h(CID_i \oplus h(PW_i)))$ and $y$, because $C_i$ is to be computed as $C_i = h(T \oplus N_i \oplus B_i \oplus y)$. It is difficult to guess the user password and very difficult to get $y$ from the user smart card. Thus, the adversary cannot forge a valid login.

3. It is hard for any adversary to derive the user password or system secret key $x$ from $N_i$, because the one-way hash function is computationally infeasible to inverse [9]. The security of $C_i$ is also based on hash function. Moreover, an adversary cannot perform an off-line guessing attack of $C_i$, because it is computed with two secret parameters $B_i$ and $y$.

4. In practice, it is likely that the user $U_i$ uses the same password $PW_i$ to access several servers for his convenience. If an insider of a remote system obtains $PW_i$, he could impersonate $U_i$'s login to access other servers. In our scheme, the user $U_i$ will not reveal $PW_i$ to the remote system, instead, the verifier obtains $h(PW_i)$ in the verification process. But, it is difficult to get $PW_i$ from its hash value. Moreover, the remote system does not maintain any verifier table through which a dishonest party can steal the user password. Thus, the proposed scheme can resist the insider attack and stolen verifier attacks.

5. If the adversary gets user password, he still cannot forge user login. To create a forged login, he also needs the secret number $y$ stored in a user smart card, which is extremely difficult to get it. Additionally, once the user comes to know that his password is leaked, the user can easily invoke the password change protocol and changes his password without any assistance of the remote systems.

Thus, the proposed scheme can resist the reply attacks, forgery attacks, guessing attacks, insider attacks and stolen verifier attacks.

## IV. CONCLUSION

We have proposed a dynamic ID-based remote user authentication scheme, which allows the users to choose and change their passwords freely. By employing a dynamic ID, the scheme is protected from ID-theft. As the scheme does not maintain any verifier table to verify the validity of the user login, it eliminates the remote system's overheads and strengthens the protocol against stolen verifier attacks. The security of the scheme is based on one-way hash function, which is infeasible to inverse. We have shown that the scheme can resist the reply attacks, forgery attacks, guessing attacks and insider attacks.

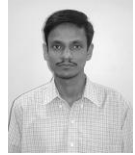

**Manik Lal Das** received his M. Tech. Degree in 1998. He is currently pursuing his Ph.D. degree in K. R. School of Information Technology, Indian Institute of Technology-Bombay, India. He is a member of Cryptology Research Society of India and Indian Society for Technical Education. His research interests include Cryptography and Information Security.

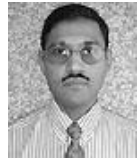

**Ashutosh Saxena** received his Ph.D. degree in Computer Science from Devi Ahilya University, Indore, India. He is an associate professor with Institute for Development and Research in Banking Technology, Hyderabad, India. He is a member of Cryptology Research Society of India and Computer Society of India. His research interests include Authentication Technologies, Smart Cards, Key Management and Security Issues in Banking.

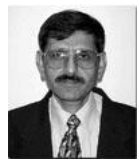

**Ved P. Gulati** received his Ph.D. degree in Computer Science from Indian Institute of Technology-Kanpur, India. He is Director of Institute for Development and Research in Banking Technology, Hyderabad, India. He is a member of IEEE, Cryptology Research Society of India and Computer Society of India. His research interests include Payment Systems, Security Technologies, Financial Networks and Banking Applications.